\documentclass[aps,prl,twocolumn,showpacs,groupedaddress]{revtex4}
\usepackage{graphicx} 
\usepackage{dcolumn}  
\usepackage{bm}       
\usepackage{amssymb}  
\usepackage{epsfig}

\def\slashchar#1{\setbox0=\hbox{$#1$}        
   \dimen0=\wd0                              
   \setbox1=\hbox{/} \dimen1=\wd1            
   \ifdim\dimen0>\dimen1                     
      \rlap{\hbox to \dimen0{\hfil/\hfil}}   
      #1                                     
   \else                                     
      \rlap{\hbox to \dimen1{\hfil$#1$\hfil}}
      /                                      
   \fi}
\def\etmiss{\slashchar{E}_T}
\def\Ptmiss{\slashchar{P}_T}

\begin{document}

\hspace{5.2in} \mbox{FERMILAB-PUB-08-295-E}

\title{\boldmath Measurement of $\sigma (p\bar{p}\to Z + X)\cdot{\rm Br}(Z\to\tau^+\tau^-)$ at $\sqrt{s} = 1.96$~TeV}

%
\author{V.M.~Abazov$^{36}$}
\author{B.~Abbott$^{75}$}
\author{M.~Abolins$^{65}$}
\author{B.S.~Acharya$^{29}$}
\author{M.~Adams$^{51}$}
\author{T.~Adams$^{49}$}
\author{E.~Aguilo$^{6}$}
\author{M.~Ahsan$^{59}$}
\author{G.D.~Alexeev$^{36}$}
\author{G.~Alkhazov$^{40}$}
% ...

\affiliation{\vspace{0.1 in}(The D\O\ Collaboration)\vspace{0.1 in}}
% ...
\affiliation{$^{6}$University of Alberta, Edmonton, Alberta, Canada,
                Simon Fraser University, Burnaby, British Columbia, Canada,
                York University, Toronto, Ontario, Canada, and
                McGill University, Montreal, Quebec, Canada}
% ...
\affiliation{$^{29}$Tata Institute of Fundamental Research, Mumbai, India}
% ...
\affiliation{$^{36}$Joint Institute for Nuclear Research, Dubna, Russia}
% ...
\affiliation{$^{40}$Petersburg Nuclear Physics Institute,
                St. Petersburg, Russia}
% ...
\affiliation{$^{49}$Florida State University, Tallahassee, Florida 32306, USA}
% ...
\affiliation{$^{51}$University of Illinois at Chicago,
                Chicago, Illinois 60607, USA}
% ...
\affiliation{$^{59}$Kansas State University, Manhattan, Kansas 66506, USA}
% ...
\affiliation{$^{65}$Michigan State University,
                East Lansing, Michigan 48824, USA}
% ...
\affiliation{$^{75}$University of Oklahoma, Norman, Oklahoma 73019, USA}
% ...

\date{August 8, 2008}

\begin{abstract}
We present a measurement of the cross section for $Z$ boson production
times the branching fraction to tau lepton pairs $\sigma (p\bar{p} \to
Z + X) \cdot$Br$(Z\rightarrow \tau^+ \tau^-)$ in $p \bar{p}$
collisions at $\sqrt{s} = 1.96$ TeV.  The measurement is performed in
the channel in which one tau lepton decays into a muon and neutrinos,
and the other tau lepton decays hadronically or into an electron and
neutrinos.  The data sample corresponds to an integrated luminosity of
1.0~fb$^{-1}$ collected with the D0~detector at the Fermilab Tevatron
Collider.  The sample contains 1511 candidate events with an estimated
20\% background from jets or muons misidentified as tau leptons.  We
obtain $\sigma \cdot \rm{Br}$ $= 240 \pm
8$\thinspace(stat)\thinspace$\pm$12\thinspace(sys)\thinspace$\pm$15\thinspace(lum)~pb,
which is consistent with the standard model prediction.
\end{abstract}


\maketitle

The resonant production of tau lepton pairs is as interesting for the
study of standard model (SM) physics as the production of lighter
lepton pairs. For new phenomena, especially for decays of particles
coupled to mass, such as SM or supersymmetric Higgs bosons, the
detection of resonant pairs of tau leptons becomes even more
interesting. This is due to the fact that tau leptons are much heavier
than the other leptons, increasing the chance that these new phenomena
would be observed first in this channel. Unfortunately, the detection
of tau leptons is far more difficult than that of muons or electrons.

A measurement of $\sigma(p\bar{p} \to Z + X) \cdot$Br($Z\rightarrow
\tau^+ \tau^-$) in $p \bar{p}$ collisions at $\sqrt{s} = 1.96$ TeV is
described in this Letter. The analysis is based on an event sample
containing a single isolated muon from a tau lepton decay and a tau
candidate reconstructed as a narrow jet that could be produced by a
tau lepton decaying either hadronically or into an electron and
neutrinos.  This measurement is of interest not only as a test of the SM
prediction but also because any excess over the expected $\sigma
\cdot$Br could be an indication of a source other than $Z$ bosons for
events containing tau lepton pairs, such as the Higgs boson
\cite{other_source_of_taus}. The precision of this result is
significantly improved compared to earlier publications
\cite{p14ZtautauPRD, CDFZtautau}.

The analysis presented here is based on data collected between
September 2002 and February 2006 by the D0\ experiment, corresponding
to an integrated luminosity of 1003 $\pm$ 62 pb$^{-1}$ \cite{newlumi}.

The D0\ detector~\cite{run2det} is a general purpose, axially and
forward-backward symetric detector, consisting of a central-tracking
system located within a 2~T superconducting solenoidal magnet,
surrounded by three liquid-argon/uranium calorimeters and a muon
detector. The spatial coordinates of the D0 detector are defined using
a righthanded Cartesian system with the origin in the center of the
detector. The positive $z$-axis is the direction of the proton beam, the
positive $y$-axis points upwards and the positive $x$-axis points out
of the Tevatron ring. The azimuthal angle $\phi$ is measured with
respect to the positive $x$ direction.  Pseudorapidity is defined as
$\eta = - \ln [\tan(\theta/2)]$, where the polar angle $\theta$
is measured with respect to the positive $z$ direction. The tracking
system has coverage up to $\eta \approx 3$. The calorimeter consists
of a central section (CC) covering $|\eta| \lesssim 1.1$ and two end
calorimeters (EC) that extend coverage to $|\eta|\approx 4.2$, all
housed in separate cryostats and segmented into cells of dimensions 
$0.1 \times 0.1$ in $\eta - \phi$ space ~\cite{run1det}. The muon
system~\cite{run2muon} provides a coverage up to $\eta \approx 2$ and
is located outside the calorimeter; it consists of a layer of tracking
detectors and scintillation trigger counters before 1.8~T iron
toroids, followed by two similar layers after the toroids. Luminosity
is measured using plastic scintillator arrays located in front of the
EC cryostats, covering $2.7<|\eta|<4.4$. A three level trigger
system is designed to select most interesting events based on
preliminary information from the tracking, calorimetry, and muon
systems, reducing the number of recorded events from the collision rate 
of $\approx$ 2~MHz to a rate of $\approx$ 50~Hz, which is written to
tape.

The triggering strategy used in this analysis is based on the tau
lepton which decays into $\mu\nu_\mu\nu_\tau$. A single muon trigger
requiring hits in the muon system in combination with a high
transverse momentum ($p_T$) track reconstructed in the central
tracking system is required. The average trigger efficiency, ultimately
parametrized as a function of $\phi$, $\eta$ and $z$ using a data sample
of $Z \rightarrow \mu^+ \mu^-$ events, is ($52.3 \pm 1.4$)\%. No dependence
on the muon $p_T$ is observed above 15~GeV.

Most backgrounds as well as the efficiency of the selection for
signal $Z \rightarrow \tau^+ \tau^-$ events are estimated using Monte
Carlo (MC) simulations.  All simulated samples are generated with {\sc
  pythia} \cite{pythia} using the CTEQ6.1L parton distribution
function (PDF) set.  Simulation of the D0\ detector is done using {\sc
  geant3} \cite{GEANT}.   Noise in the
detector and the contributions from other simultaneous interactions
are simulated by adding random untriggered data events to the MC
simulation. These events were chosen such that the effective
instantaneous luminosity distribution in MC is the same as in data.
The code used for the reconstruction of simulated events is identical
to the one used for data.

Corrections are applied to all MC events to obtain overall good
agreement between the simulation and collider data. The momentum scale
and resolution for muons in the MC are tuned to reproduce the $Z$
boson invariant mass distribution observed in data. Similarly, the jet 
energy resolution is tuned to match that observed in data for each region 
of the detector. The $p_T$ spectrum of the $Z$ boson for events generated
with {\sc pythia} has a different shape than that measured in data;
therefore the $p_T$ of the $Z$ boson is reweighted to fit the direct
measurement in data \cite{ZpTpaper}. Small differences in acceptance
between data and simulation are corrected for by weighting the
simulated $z$ position of the primary vertex in MC events to reproduce
that observed in data.

\begin{figure*}
\epsfig{figure=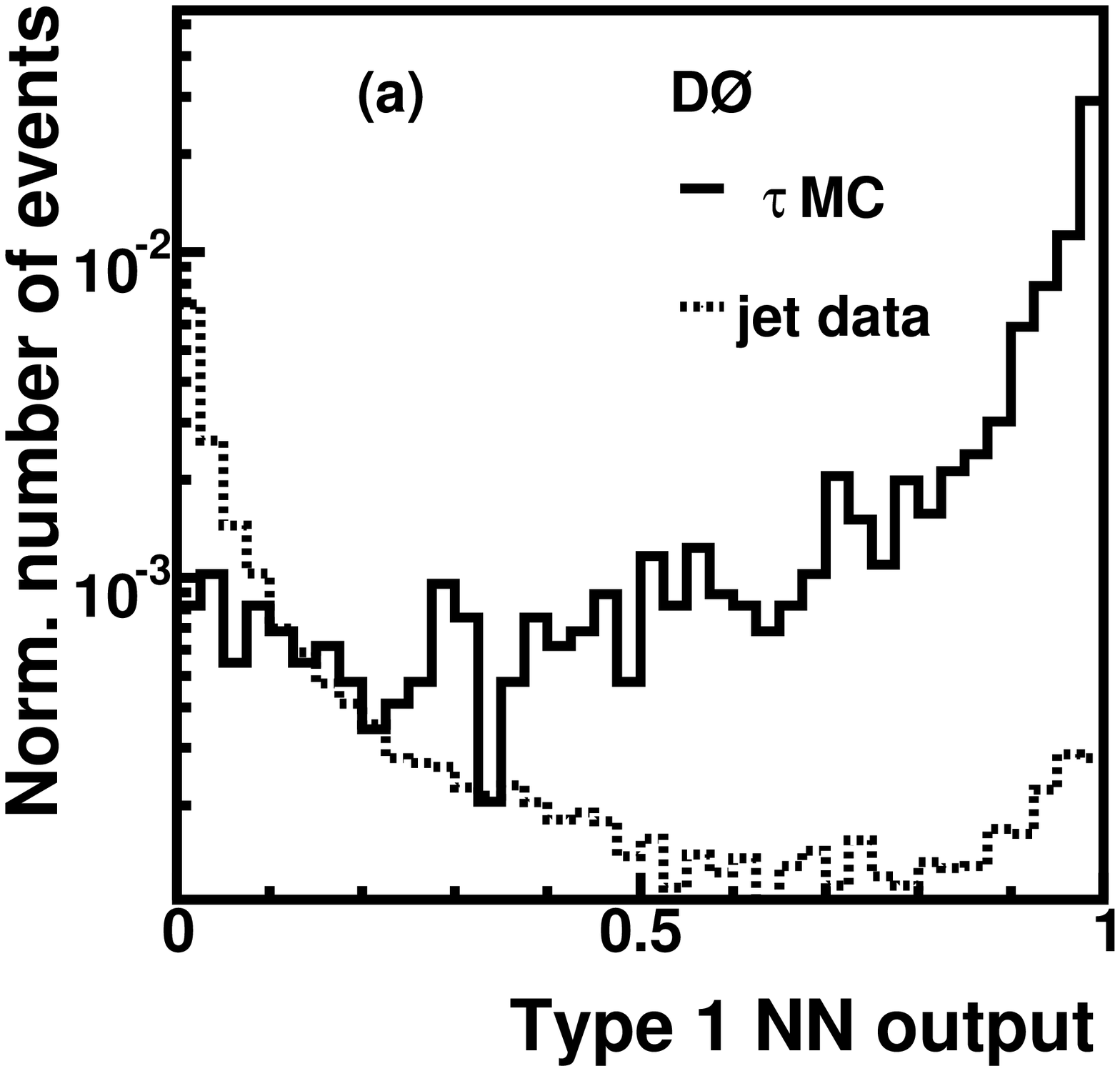,height=5.2cm,width=5.9cm}
\epsfig{figure=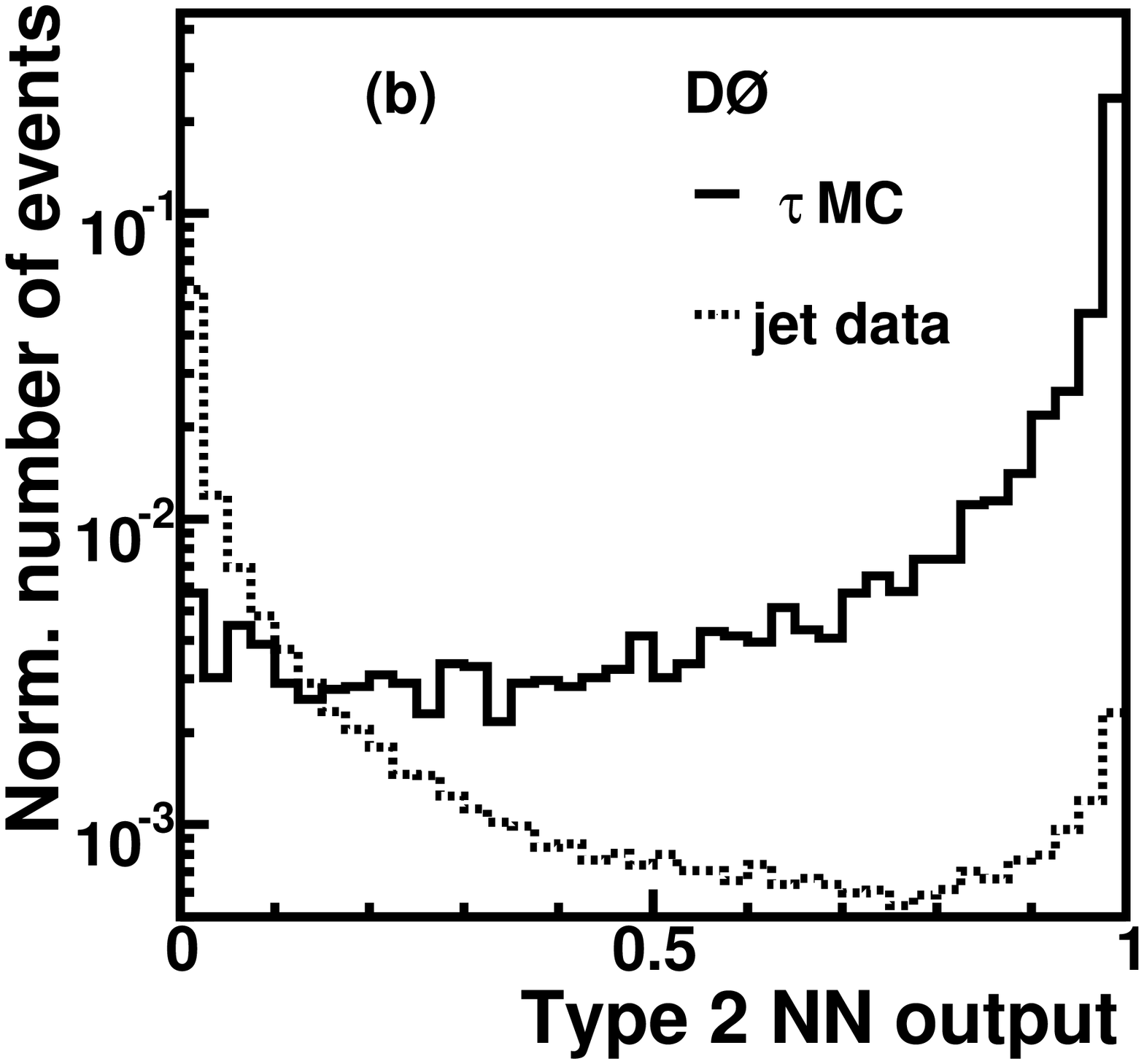,height=5.2cm,width=5.9cm}
\epsfig{figure=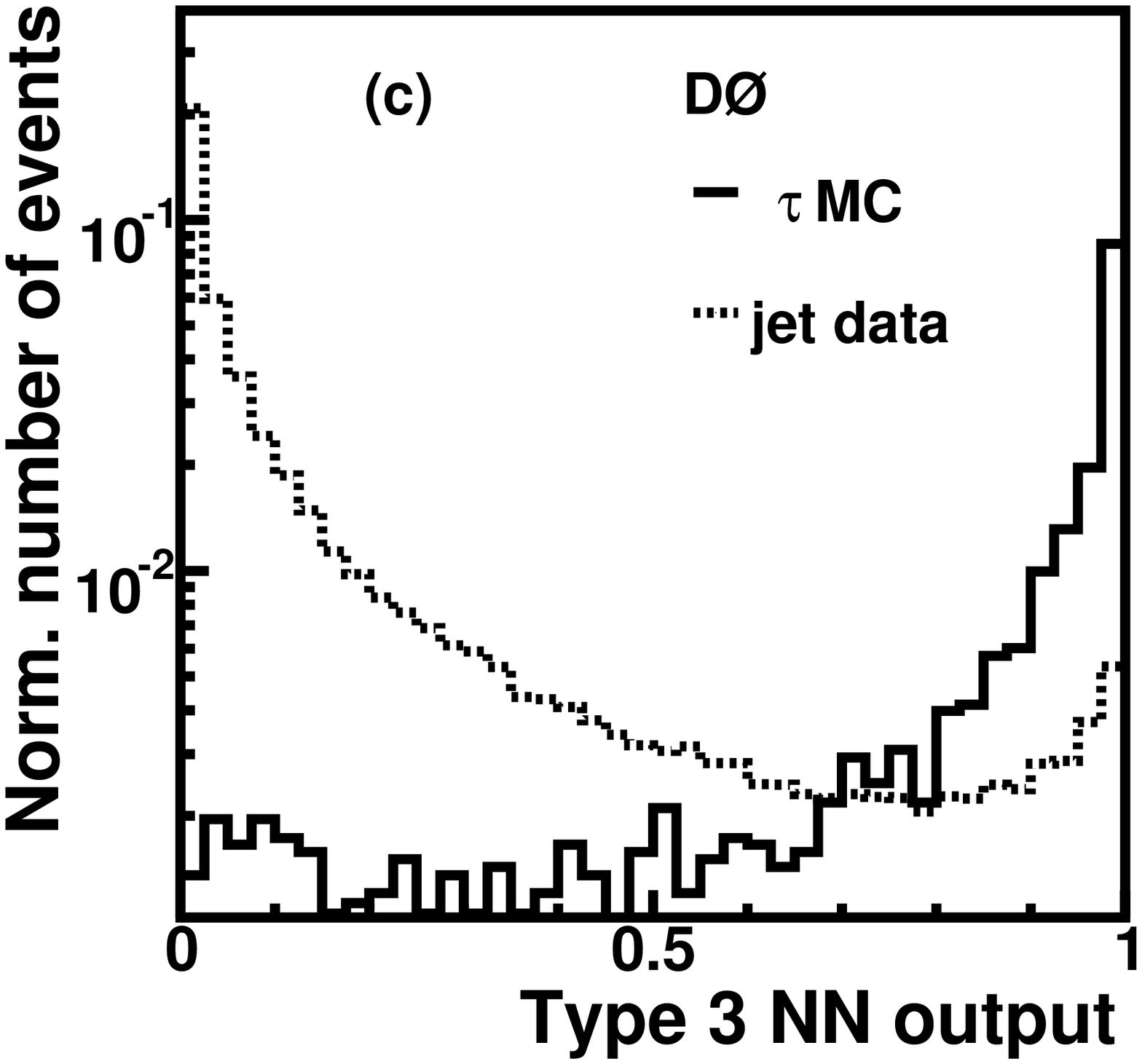,height=5.2cm,width=5.9cm}
\caption{\label{fig:NN} NN output distributions for (a) type 1, (b)
  type 2, and (c) type 3 tau candidates. The ratio of signal to
  background is arbitrary, but the relative amounts of type 1, type 2,
  and type 3 events in background and signal are not. The
  distributions are normalized with respect to each other such that
  the sum over the three types is 1 for both signal and background.}
\end{figure*}

Reconstruction efficiencies for muons and tracks are calculated both
in data and MC using samples of $Z \rightarrow \mu^+ \mu^-$
events. Efficiency correction factors for MC events as a function of
muon or track $\phi$, $\eta$ and $z$ are applied.  The signal or
background samples are normalized to the expected number of events
evaluated using the luminosity of the data sample and the theoretical
values of the next-to-next-to-leading order (NNLO) cross sections in
the case of $Z$ boson production
\cite{NNLO_Zcrosssection1,NNLO_Zcrosssection2} or next-to-leading
order (NLO) cross sections for all other processes where the NNLO
calculation is not available. $W$ boson production is
normalized from data.

In this analysis, muons are identified starting with their signature
in the muon detector.  The track reconstructed from hits in the muon
layers is required to match a track from the central tracking
detectors. The muon momentum is measured using only the central
tracking detectors.

A tau candidate is a collection of (i) a calorimeter cluster
reconstructed using a simple cone algorithm \cite{simple_cone}, (ii)
tracks associated with the calorimeter cluster of which at least one
has $p_T>1.5$~GeV but with a total invariant mass less than $1.8$~GeV,
and (iii) electromagnetic (EM) sub-clusters constructed from the cells
in the EM section of the calorimeter. The size of the cone used for
reconstruction of the calorimeter cluster is ${\cal R}=\sqrt{(\Delta
  \phi)^2 + (\Delta \eta)^2}=0.5$, where $\Delta \phi$ is the
difference in azimuthal angle, and $\Delta \eta$ the difference in
pseudorapidity between the cone axis and each of the calorimeter
towers. Isolation variables are calculated using a cone of ${\cal
  R}=0.3$.  The tracks associated with the tau candidate must also be
contained within this ${\cal R}=0.3$ cone.

Tau candidates are classified as type 1, 2 or 3, depending on the
numbers of tracks and EM sub-clusters they possess. Type 1 tau
candidates have exactly one associated track and no EM sub-clusters,
type 2 have one associated track and one or more EM sub-clusters, and
type 3 have at least two associated tracks.  These categories
correspond roughly to pure one-prong decays, one-prong plus neutral
pion decays as well as decay into electrons, and three-prong decays of
the tau lepton.

Due to the large number of jets reconstructed as tau candidates,
additional selection criteria must be applied in order to distinguish
tau leptons from jets. Three neural networks (NN), one for each tau
type, are trained using $Z \to \tau^+ \tau^-$ MC events as signal and
events with a jet back-to-back with a non-isolated muon from data as
background. The NNs use isolation variables based on tracks, hadronic
and EM calorimeter clusters, as well as shower shape variables and
correlation variables between calorimeter and
tracks. Figure~\ref{fig:NN} shows the discrimination obtained using
the NNs. Requiring that the NN output be larger than 0.9 results in a
background rejection of almost a factor of 50 for all three tau
types. This reduces the probability for a jet to be misidentified as a
tau lepton to 1.1\% for the sum of all types (from 52\% without the NN
output requirement), while maintaining a total efficiency of close to
70\% for tau leptons which decay hadronically or to an electron and
neutrinos. Electrons are treated as type 2 tau candidates, since the
efficiency for them to be reconstructed as such and pass the NN output
requirement is 98\%. For a complete description of the neural networks
and details on their performance see Ref.~\cite{confprocCristina}.

The variable chosen to best illustrate the $Z \to \tau^+\tau^-$ signal
is the visible mass, given by:
\begin{equation}
{\rm Visible~Mass  = \sqrt{( P_{\mu} +  P_{\tau} + \Ptmiss )^2} },
\end{equation}
where ${\rm P}_{\mu,\tau}=(E_{\mu,\tau},p_{\mu,\tau}^x,
p_{\mu,\tau}^y, p_{\mu,\tau} ^z)$ are the four-momentum vectors of the
muon and the tau candidate, and $\Ptmiss=(\etmiss, \etmiss^x,
\etmiss^y, 0)$, with $\etmiss$ being the missing transverse energy in
the event and $\etmiss^x, \etmiss^y$ being its projections along the
$x$ and $y$ directions. The uncorrected missing transverse energy is 
defined as the vector equal in length and opposite in direction to the
vectorial sum of transverse energies of the calorimeter cells. 
The transverse momenta of muons are subtracted from this vector, after
corrections for the energy deposited by the muons in the calorimeter
have been applied.  When the tau candidate matches a reconstructed
electron, the energy corrections derived for electrons are applied.
For jets corresponding to tau candidates, the tau energy corrections
described below are applied. Jet energy corrections applied to all
other jets in the event are propagated to the missing $E_T$
calculation.

To compare the visible mass distributions of the tau pairs between
data and MC, it is important to have the correct energy scale for the
tau candidate. For type 1 tau candidates, the momentum of the track is
used as the best estimate of the energy of the tau candidate when the
tracking resolution is superior to the calorimeter energy resolution
(up to calorimeter cluster energy of 70~GeV). For type 2 candidates
matching electrons, the energy corrections derived for electrons are
applied.  For type 2 candidates not matching electrons and type 3 tau
candidates, the energy is estimated using
\begin{equation}
 E^{\text{corr}} = \sum_{i} p_i^{\text{trk}} + E^{\text{cal}} -
 \sum_{i} R(p_i^{\text{trk}},\eta)\cdot p_i^{\text{trk}},
\end{equation}
where $p_i^{\text{trk}}$ is the momentum of track $i$ associated with
the tau candidate, $E^{\text{cal}}$ is the energy deposited by the tau
candidate in the calorimeter, and $R(p_i^{\text{trk}},\eta)$
represents the response of the calorimeter to the $\pi^\pm$ which
produced track $i$ associated with the tau candidate, as a function of
the energy and rapidity of the $\pi^\pm$. Typically, $0.6 <
R(p_i^{\text trk}, \eta) < 0.9$.  As the resolution of the calorimeter
is better than that of the tracking at calorimeter cluster energies
higher than 70~GeV (type 1), 100~GeV (type~2), or 120~GeV (type~3),
the energy of the calorimeter cluster is used in these cases, after
applying $\eta$ and energy dependent corrections obtained from MC.

The default program in the D0 {\sc geant} simulation for hadronic
interactions, {\sc geisha} \cite{Geisha}, does not reproduce the
charged pion response in data well.  Therefore g{\sc calor}
\cite{gCALOR} is used for a more precise simulation of single charged
pion interactions.  The charged pion response obtained using these
special simulations was found to be in reasonable agreement with
preliminary data measurements in the central calorimeter
\cite{thesis_Cristina}. The energy measurement for neutral particles,
mostly important for type~2 taus, is dominated by electromagnetic
showers in the calorimeter. The simulation of electromagnetic showers 
in {\sc geant} is sufficiently accurate for the purpose of this
measurement.

The preselection requires one isolated muon reconstructed within the
pseudorapidity interval $ | \eta| < 1.6$. The transverse momentum of
the muon as measured by the central tracking detectors must satisfy
$p_T^\mu > 15$~GeV. No other muon matched to a central track with $p_T
> 10$~GeV is allowed in the event.  The muon isolation requires the
sum of energies of all cells situated in a hollow cone around the
direction of the muon with $0.1<{\cal R}<0.4$, as well as the sum of
all tracks in a cone of ${\cal R}<0.5$, excluding the muon track, to
be less than~2.5~GeV.

The preselection further requires one tau candidate with $p_T >
15$~GeV, $|\eta| < 2$, scalar sum of the transverse momenta of all
tracks associated with the tau candidate $>15$~GeV for types 1 and 3
and $> 5$~GeV for type 2 tau candidates, $NN > 0.3$, and no other muon
matching the tau candidate.  Type 3 tau candidates with two tracks are
only considered if both tracks have the same charge. The tau candidate
is required to have a charge with opposite sign to that of the muon.
The distances in the $z$ direction at the track's point of closest
approach between the muon and the primary vertex, the tau candidate
and the primary vertex, as well as the distance between the muon and
the tau candidate must be less than 1 cm.

In total 8316 events pass these criteria.  To reduce the $W$ + jets
and the $Z \to \mu^+ \mu^-$ backgrounds, another selection criterion
is used, based on a variable which gives an approximation of the $W$
boson mass, referred to as $m^*$:
\begin{equation}
m^*  = \sqrt{2 E_{\nu} E_{\mu} (1 - \cos \Delta \phi)},
\end{equation}
where $E_{\nu} = \etmiss \cdot E_{\mu}/p_T^\mu$ is an approximation of
the neutrino energy, and $\Delta \phi$ is the angle between $\etmiss$
and the muon in the transverse plane.

For the final selection, all the preselection criteria are
applied. Additionally, the lower limit on the NN output for the tau
candidates is raised to 0.9 for types 1 and 2, and to 0.95 for type 3
tau candidates.  The final selection also requires $m^* < 20$~GeV.  A
total of 1511 events pass all the selection criteria.

The dominant remaining background arises from multijet processes,
mainly from $ b \bar{b}$ events where the muon isolation requirement
is met and one of the jets satisfies the tau candidate selection
criteria. Another significant source of events with isolated muons and
tau candidates is $W$ + jets production, where the $W$ boson decays to
$\mu\nu$ and one of the jets is misidentified as a tau candidate.  The
$Z \rightarrow \mu^+ \mu^-$ background is reduced by the requirement
that no other muon be found in the event, but a small number of events
will be selected when one of the muons is not reconstructed.  Small
contributions are also expected from $W \rightarrow \tau \nu$ and $WW
\rightarrow l \nu l \nu$, as well as $t \bar t$ production.
Contributions from $WZ$ and $ZZ$ events yield less than one event each
after the final selection criteria and are therefore neglected.  All
backgrounds, except the multijet background, are estimated using MC
simulations.

The multijet background is estimated using the data events that
satisfy all requirements placed on the signal sample except that the
muon and the tau candidate have the same sign charge.  We call this
the same-sign (SS) sample.  To obtain the appropriate normalization
for this background, a special data sample is selected, the ``multijet
sample,'' containing events that pass all requirements placed on the
signal sample except the isolation criteria and the cut on the tau NN
output.  Instead of the isolation requirement used for the signal
events, the events in the multijet sample have the sum of energies of
all calorimeter cells inside the hollow isolation cone in the range
2.5 to 10~GeV. The sum of all non-muon tracks $p_T$ within the track
isolation cone is required to be in the same interval $2.5 - 10$~GeV.
To avoid contamination from $Z \rightarrow \tau^+ \tau^-$ signal
events, an upper limit on the tau NN output is placed at 0.8.  To
increase the statistics of this sample, the muon $p_T$ is required to
be at least 10~GeV instead of 15~GeV.  The multijet sample is expected
to be completely dominated by multijet processes, but may also include
events in which a $W$ decaying into a muon is produced in association
with a jet. The $W + {\rm jets}$ contribution is reduced by requiring
that the muon and the tau candidate are back to back ($|\phi_\mu -
\phi_\tau| > 2.5$).  A slight excess of opposite sign (OS) over SS 
events is observed in the multijet sample.  No significant dependence 
of the OS/SS ratio as a function of $p_T$ or NN output is observed for 
the three types of tau candidates in the multijet sample.  Correction 
factors ($f_{\rm  mj}^i$) of $1.13 \pm 0.03, 1.08 \pm 0.01$, and 
$1.06 \pm 0.01$ for tau types 1 to 3 are obtained, being used as discussed 
below to normalize the multijet background in the final signal sample.

The number of events in the SS sample is corrected for the
contribution from $Z \rightarrow \mu^+ \mu^-$, $Z \rightarrow \tau^+
\tau^-$, and $W \rightarrow \tau \nu$ obtained from MC, multiplied by
an additional correction factor which takes into account the
difference between the charge misidentification rates in data and
MC. Totals of 6 events for type 1, 16 events for type 2, and 18 events 
for type 3 tau candidates from $Z \rightarrow \mu^+ \mu^-$, 
$Z \rightarrow \tau^+\tau^-$, and $W \rightarrow \tau \nu$ are estimated 
to have a misidentified charge after all cuts and are subtracted from the 
number of SS events when the multijet background is calculated. The
contribution from $W \rightarrow \mu \nu$ events is accounted for
separately.

A part of the $W$ + jets background is already included in the SS
sample.  However, we expect a significant excess of OS events compared
to the number of SS events due to the fact that a high percentage of
$W + 1~\rm{jet}$ events comes from quark jets.  The number of $W +
\rm{jets}$ events in data is estimated by selecting a sample that is
expected to have a large contribution from $W$ boson processes and low
or negligible contributions from $Z$ boson production.  Such a
$W$+jets enriched sample can be obtained by requiring an isolated muon
with $p_T > 20$~GeV, a tau candidate with 0.3 $< NN < $ 0.8,
$|\phi_\mu - \phi_\tau| < 2.7$, and $m^* > 40$~GeV. Mostly multijet
and $W$+jets events contribute to this sample. The excess of OS events
compared to SS events is given for the multijet background by $f_{\rm
  mj}^i$ for tau type~$i$. For the $W + {\rm jets}$ sample, similar
factors ($f_W^i$) of 2.39 $\pm$ 1.01, 3.15 $\pm$ 1.17, and 1.6 $\pm$
0.26 are estimated from data, in the sample with the cuts listed
above, but requiring a tighter cut $m^* > 60$~GeV. Using this, we can
calculate the number of $W + {\rm jets}$ events in the $W + {\rm
  jets}$ enriched data sample by solving the following system of two
equations for each tau type $i$:
\begin{equation}
N_{W}^i + N_{\rm mj}^i = N_{OS}^i + N_{SS}^i 
\end{equation}
\begin{equation}
\frac{f_{W}^i - 1}{f_{W}^i +1}N_{W}^i +\frac{f_{\rm mj}^i -
  1}{f_{\rm mj}^i +1} N_{\rm mj}^i = N_{OS}^i - N_{SS}^i
\end{equation}
where $N_{W}^i$ is the number of $W + {\rm jets}$ events, $N_{\rm mj}^i$
is the number of multijet events and $N_{OS}^i$, $N_{SS}^i$ are the numbers
of OS, respectively SS events in the  $W + {\rm jets}$ enriched data sample. 
The ratios between the number of $W$ + jets events calculated in data by 
solving the above system of equations and the one expected from MC
for each tau type are used as normalization factors for this
background in the signal region.  The uncertainty on $N_{W}^i$ from
data is taken as a systematic uncertainty.  The estimated number of
$W$ + jets events in the signal sample, not including those in the SS
sample, is $14 ~\pm$ 5 events.
\begin{figure*}
\epsfig{figure=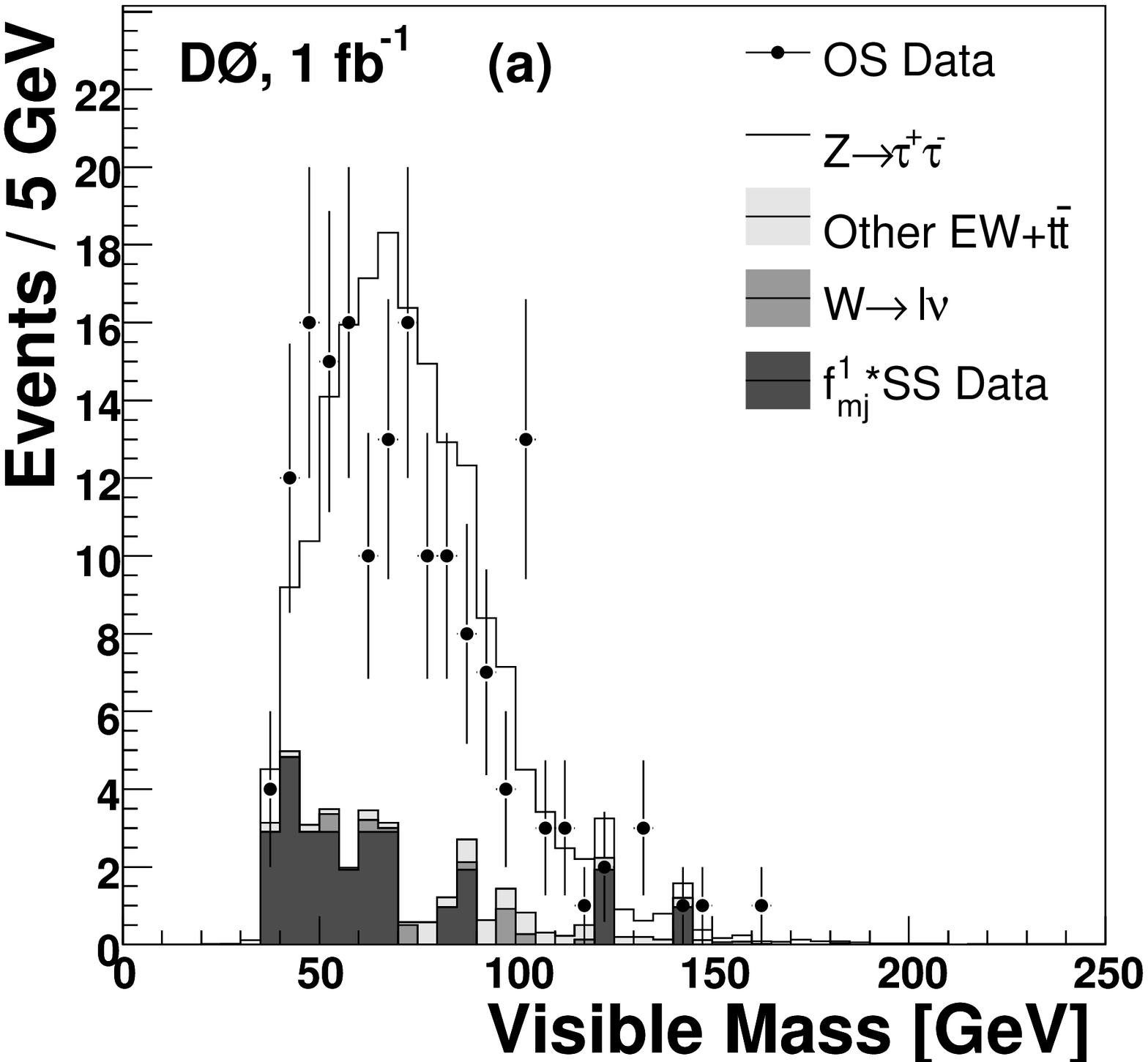, width=5.911cm, height=5.2cm} 
\epsfig{figure=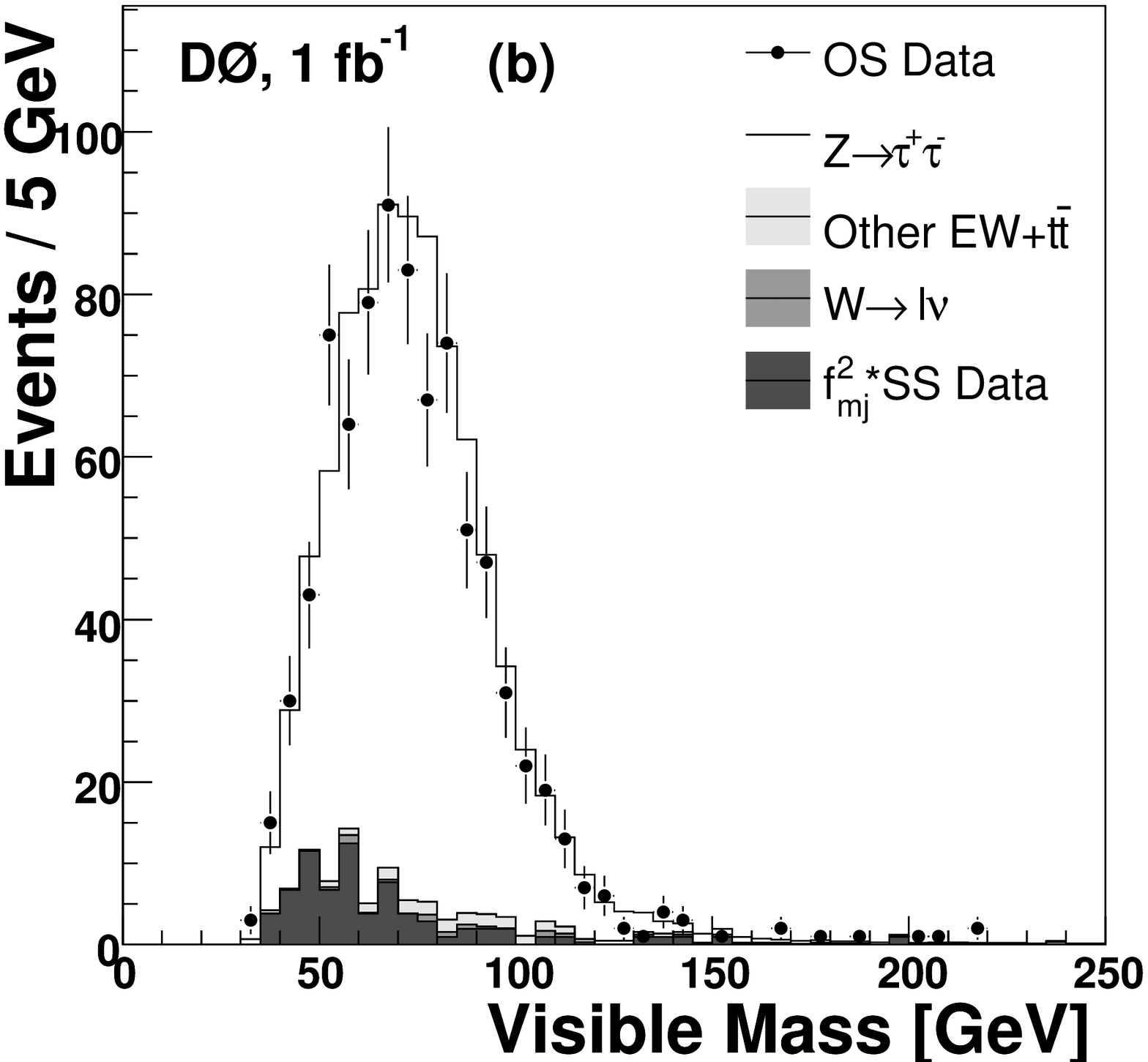,  width=5.911cm, height=5.2cm}
\epsfig{figure=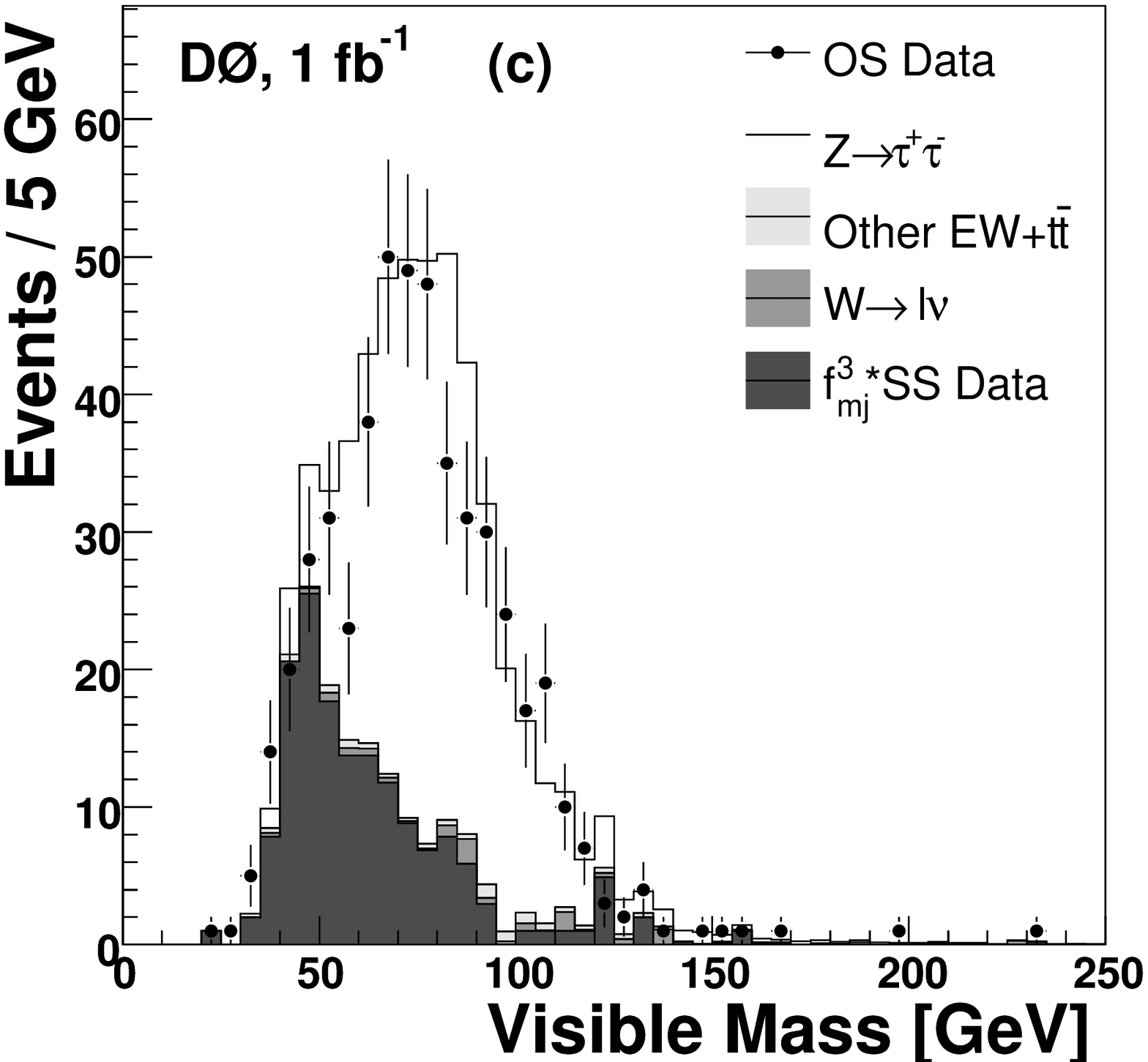, width=5.911cm, height=5.2cm}

\caption{Visible mass distribution for (a) type 1 tau events, (b) type
  2 tau events, and (c) type 3 tau events.  The data are the points
  with error bars. The different components of the SM expectation are
  as given in the legend. The $Z \to \tau^+ \tau^-$ signal is
  normalized to the theoretical expectation calculated at NNLO using
  MRST2004 PDFs \cite{NNLO_Zcrosssection1,NNLO_Zcrosssection2}.
\label{fig:mvis}}
\end{figure*}

Several distributions such as muon and tau candidate transverse
momentum, pseudorapidity, and azimuthal angle, as well as $\etmiss$,
$m^*$, and visible mass are compared between the data and the
predicted sum of backgrounds and $Z \rightarrow\tau^+ \tau^-$ for the
SM cross section and branching ratio. All these distributions show
good agreement after each of the preselection, NN selection, and
anti-$W$ requirement stages.

\begin{figure}
\vspace{-0.7cm}
\epsfig{figure=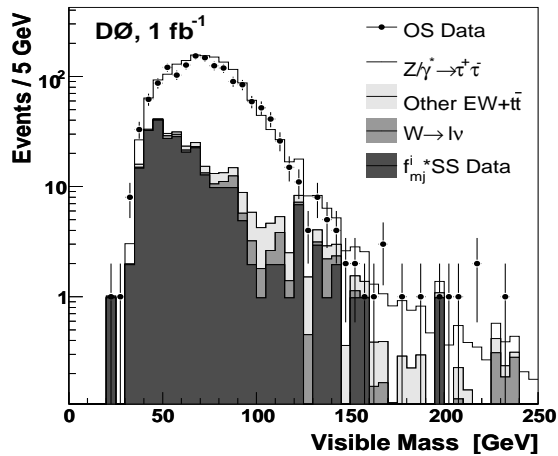, width=7.8cm, height =6.6cm}
\caption{\label{fig:mvislog} Visible mass distribution for all tau
  types. The $Z \to \tau^+ \tau^-$ signal is
  normalized to the theoretical expectation calculated at NNLO using
  MRST2004 PDFs.}
\end{figure}

In Fig.~\ref{fig:mvis} the visible mass distribution for events which
pass the final selection requirements is shown separately for each of
the tau types, while Fig.~\ref{fig:mvislog} shows the same
distribution for the sum of all types.  Good agreement is observed
between the data and the sum of the background SM processes and $Z
\rightarrow\tau^+\tau^-$ signal, normalized using the NNLO SM
prediction \cite{NNLO_Zcrosssection1,NNLO_Zcrosssection2}.

Table \ref{tab:events} shows the number of events expected for each
tau type from each of the backgrounds, as well as from the $Z
\rightarrow \tau^+ \tau^-$ signal.  It also shows the total numbers of
expected background and signal events in comparison to the numbers of
events observed in data for three levels of selection: preselection,
preselection plus NN output requirement, and after all selection criteria
are applied. Good agreement is observed between the predicted and
observed numbers of events at each level of selection for all tau
types.

We estimate that approximately 1.2\% of all $Z \rightarrow \tau^+
\tau^-$ events have the wrong sign for either the muon or the tau
candidate, therefore appearing as SS events.  From the number of $Z
\rightarrow \tau^+ \tau^-$ events obtained by subtracting the
estimated background from the number of events in the final sample, we
calculate the number of $Z \rightarrow \tau^+ \tau^-$ events
reconstructed as SS to be 17. This number is added to the number of
events in the OS sample when calculating the $Z \rightarrow \tau^+
\tau^-$ cross section.

Reconstruction of a second track close to a first reconstructed track
is found to be more efficient in MC than in data.  A correction factor
of 0.97 $\pm$ 0.028 is applied to simulated events containing type 3
tau candidates. This factor is obtained by comparing the ratios of
type 3 tau candidates with two and three tracks in data and MC and
taking into account that there are twice as many SS as OS combinations
when one of the three tracks is lost.

\begin{center}
\footnotesize
\begin{table*}
\begin{center}
\begin{ruledtabular}
\begin{tabular}{l|r@{$\,\pm \,$}lr@{$\,\pm \,$}lr@{$\,\pm \,$}l|r@{$\,\pm \,$}lr@{$\,\pm \,$}lr@{$\,\pm \,$}l|r@{$\,\pm \,$}lr@{$\,\pm \,$}lr@{$\,\pm \,$}l}
 & \multicolumn{6}{c|}{Type 1}  & \multicolumn{6}{c|}{Type 2}  & \multicolumn{6}{c}{Type 3} \\
\hline
Process&\multicolumn{2}{c}{Preselection}&\multicolumn{2}{c}{Preselection}&\multicolumn{2}{c|}{All cuts}&\multicolumn{2}{c}{Preselection}&\multicolumn{2}{c}{Preselection}&\multicolumn{2}{c|}{All cuts}&\multicolumn{2}{c}{Preselection}&\multicolumn{2}{c}{Preselection}&\multicolumn{2}{c}{All cuts}\\
       &   \multicolumn{2}{c}{}           &\multicolumn{2}{c}{+ NN $>$ 0.9}   & \multicolumn{2}{c|}{}   &     \multicolumn{2}{c}{}           &      \multicolumn{2}{c}{+ NN $>$ 0.9}    &    \multicolumn{2}{c|}{}           &     \multicolumn{2}{c}{}     &    \multicolumn{2}{c}{+ NN $>$ 0.95}    &\multicolumn{2}{c}{}   \\ 
\hline
$Z /\gamma^*\rightarrow \tau^+ \tau^-$ & 302& 4 & 230 & 4  & 146 & 3  & 1469 & 9  & 1131 & 8  & 786 & 7 & 693 & 6  & 484 & 5  & 358 & 5 \\
$Z/\gamma^* \rightarrow \mu^+ \mu^-$ &  58 & 2 & 43 & 2   & 6.1 & 0.6& 176 & 3  &  108 & 3 & 14.0 & 0.8 & 184 & 3 & 38 & 1  & 8.9 & 0.7 \\
$WW$ & 7.2 & 0.3 & 6.1 & 0.3 & 0.4 & 0.1 & 79 & 1 & 74 & 1 & 6.9 & 0.3& 9.3 & 0.4 & 6.1 & 0.5 & 0.5 & 0.1\\
$t\bar{t}$ &2.7 & 0.3 & 2.0 & 0.3 & 0.2 & 0.1 & 33 & 1 & 28 & 1 & 2.4 & 0.3 & 29 & 1 & 4.2 & 0.4 & 0.5 & 0.1\\
$W \rightarrow \tau \nu$ & 10 & 2 & 4 & 1 & 1.5 & 0.8& 50 & 4 & 14.1 & 2.2 & 1.4 & 0.7& 168 & 7 & 22.0 & 2.7 & 3.7 & 1.2\\
$W \rightarrow \mu \nu$ & 127 & 11 & 42 & 5   & 2.1 & 0.9 & 470 & 18 & 116 & 9  & 6.7 & 1.9& 1384 & 32 & 202 & 13  & 14.1 & 2.7 \\
Multijet & 208 &15 & 46 & 8 &  25 & 5 & 584 & 25 & 123 & 12  & 61 & 8& 2265 & 47 & 273 & 18 & 145 & 13\\
\hline
Predicted & 715 & 18 & 373 & 11 & 181 & 7 & 2861 & 32 & 1594 & 18 & 878 & 12& 4732 & 59 & 1029 & 23 & 531 & 15 \\
\hline
Data & \multicolumn{2}{c}{720~~~~~} & \multicolumn{2}{c}{380~~~~~} & \multicolumn{2}{c|}{170}  & \multicolumn{2}{c}{2836~~}  & \multicolumn{2}{c}{1546~~}  & \multicolumn{2}{c|}{843}  & \multicolumn{2}{c}{4760~~} & \multicolumn{2}{c}{981~~~}  & \multicolumn{2}{c}{498} \\
\end{tabular}
\end{ruledtabular}

\caption{\label{tab:events} Number of OS events expected for each tau
  type from the $Z \rightarrow \tau^+ \tau^-$ signal as well as from
  each of the backgrounds, their sum and the number of OS events
  observed in data, for three levels of selection: preselection,
  preselection + NN output $>$ 0.9 (0.95 for type 3) and after all
  selection criteria are applied (preselection + NN output $>$ 0.9 or
  0.95 + $m^* < 20$~GeV). The uncertainies are statistical. }
\end{center}
\end{table*}
\end{center}

\vspace{-0.5cm} 
Systematic uncertainties on the multijet and $W + {\rm
  jets}$ backgrounds are derived from the statistical uncertainties of
the control samples used to estimate these backgrounds and from the
systematic uncertainties on the correction factors used for their
normalization.

The systematic uncertainty related to the tau energy measurement is
estimated by scaling the charged pion response used for data by the
largest difference found between the response measured in data and the
response obtained using g{\sc calor} (6\%) and recalculating the
acceptance applying all cuts. The value of this uncertainty is~1\%.

NN systematic uncertainties are calculated using statistical ensembles
of events in which each input variable is allowed to fluctuate within
the difference observed between the distributions of that particular
variable in data and MC.  The RMS of the ratio of the number of events
passing a certain NN cut to the number of events in the ensembles,
called the ensemble cut ratio, is taken as a measure of the NN
uncertainty.  The estimated uncertainties are 4.3\% for type~1, 2.0\%
for type 2, and 3.8\% for type 3 tau candidates, which results in a
total uncertainty of 2.7\%.

The uncertainty due to the tau candidate track reconstruction
efficiency is taken to be the same as the uncertainty on
reconstructing muon tracks and is estimated using $Z \to \mu^+ \mu^-$
events to be 1.4\%.  The uncertainty on the correction factor due to
differences between data and MC in tracking efficiency for type 3 taus
is added in quadrature to this value, resulting in a total uncertainty
related to the tau candidate tracks of 1.6\%. The systematic
uncertainties due to muon identification and muon track matching are
determined to be 0.6\% and 0.8\%, respectively. The systematic
uncertainty due to the charge misidentification is 1\%.  The
uncertainty on the trigger efficiency is 2.7\% and takes into account
the bias related to the choice of the control sample, the variation
due to possible background contamination, variations in time or due to
changing luminosity, the choice of binning, and the choice of
parameters for the efficiency, as well as the limited statistics.  The
uncertainty on the total integrated luminosity is 6.1\%
\cite{newlumi}, with an additional systematic uncertainty of 1\%
related to the influence on the luminosity of applying the data
quality criteria used to reject events with coherent calorimeter
noise.

The PDF uncertainty of 2.0\% is estimated using a NLO calculation
\cite{Melnikov-Petriello} and the CTEQ6.1 error sets. This uncertainty
is obtained from the variation in acceptance when these error sets are
used, added in quadrature with the difference in acceptance when using
the MRST2004 error sets at NLO and with the additional variation when
going from NLO to NNLO with MRST2004. Table \ref{tab:syst} summarizes
all the systematic uncertainties.
\begin{table}[h!]
\begin{center}
\begin{ruledtabular}
\begin{tabular}{ll}
Source & Value\\
\hline
Tau energy scale & 1.0 \% \\
Tau identification  & 2.7 \% \\
Tau track reconstruction & 1.6 \% \\
Multijet background & 1.6 \% \\
W $\rightarrow \mu \nu$ background & 0.5 \% \\
Trigger & 2.7 \% \\
Muon track match & 0.8 \% \\ 
Muon identification & 0.6 \% \\
Muon momentum resolution & 0.4 \%\\
Charge misidentification & 1.0 \% \\
MC statistics & 0.6 \% \\
PDF & 2.0 \% \\
\hline
Total (except luminosity) & 5.2 \% \\
\hline
Luminosity & 6.2 \% \\
\end{tabular}
\end{ruledtabular}
\caption{\label{tab:syst} Systematic uncertainties on the $\sigma(p
  \bar{p} \rightarrow Z/\gamma^* + X) \cdot {\rm Br}(Z/\gamma^*
  \rightarrow \tau^+ \tau^-)$ measurement.}
\end{center}
\end{table}

The cross section times branching ratio for the process $p \bar{p}
\rightarrow Z/\gamma^*+X \rightarrow \tau^+ \tau^- + X$ is given by
the number of signal events divided by the product of the total
efficiency and the integrated luminosity. The number of signal events
estimated from Table~\ref{tab:events}, with the correction for signal
events reconstructed as SS, is 1227.  Since Table~\ref{tab:events}
shows the estimated number of events in the $Z/\gamma^*$ mass range
$15-500$~GeV, other corrections have to be made in order to compare
the result of this analysis with theoretical cross sections.  To limit
the mass range to $60-130$~GeV, the number of events expected from the
mass region $15-60$~GeV (7~events) as well as from the $130-500$~GeV
mass region (26 events) are subtracted from the number of signal
events in data.  The total efficiency for $Z \rightarrow \tau^+
\tau^-$ events in the $60-130$~GeV mass region is $4.9 \times
10^{-3}$, which also includes the trigger efficiency of
52.3\%. Finally, a factor of 0.98 \cite{098factor} is applied to
estimate the $Z$ boson cross section as opposed to the $Z/\gamma^*$
cross section for this mass region.

Given the systematic uncertainties listed in Table~\ref{tab:syst} and
an integrated luminosity of 1003~$\rm{pb}^{-1}$, we estimate $\sigma(p
\bar{p} \rightarrow Z + X) \cdot {\rm Br}(Z \rightarrow \tau^+
\tau^-)$ = 240 $ \pm$ 8~ (stat) $\pm$ 12~ (sys) $\pm$ 15~ (lum) pb,
which is in good agreement with the SM prediction of
$251.9^{+5.0}_{-11.8}~\rm{pb}$
\cite{NNLO_Zcrosssection1,NNLO_Zcrosssection2} that results from the
NNLO calculation using the MRST2004 PDFs, as well as with the
$241.6^{+3.6}_{-3.2}~\rm{pb}$
\cite{NNLO_Zcrosssection1,cross_sect_CTEQ} value obtained at NNLO
using the CTEQ6.1M PDF parametrization. This result is the most
precise measurement of $\sigma(p \bar{p} \rightarrow Z + X) \cdot {\rm
  Br}(Z \rightarrow \tau^+ \tau^-)$ to date, in
good agreement with previous measurements of the $Z$ boson cross
section times branching ratio to leptons at $\sqrt{s} = 1.96$~TeV
\cite{p14ZtautauPRD,CDFZtautau,measured_crosssections}.

%
We thank the staffs at Fermilab and collaborating institutions, 
and acknowledge support from the 
DOE and NSF (USA);
CEA and CNRS/IN2P3 (France);
FASI, Rosatom and RFBR (Russia);
CNPq, FAPERJ, FAPESP and FUNDUNESP (Brazil);
DAE and DST (India);
Colciencias (Colombia);
CONACyT (Mexico);
KRF and KOSEF (Korea);
CONICET and UBACyT (Argentina);
FOM (The Netherlands);
STFC (United Kingdom);
MSMT and GACR (Czech Republic);
CRC Program, CFI, NSERC and WestGrid Project (Canada);
BMBF and DFG (Germany);
SFI (Ireland);
The Swedish Research Council (Sweden);
CAS and CNSF (China);
and the
Alexander von Humboldt Foundation (Germany).

\end{document}